\documentclass[conference]{IEEEtran}
\IEEEoverridecommandlockouts
\usepackage{algorithm}
\usepackage{algorithmic}
\usepackage[mode=buildnew]{standalone}
\usepackage{amsmath,mathtools,nccmath}
\usepackage{siunitx}
\usepackage{cite}

\usepackage{dblfloatfix}
\usepackage{booktabs}
\usepackage{multirow}

\usepackage{enumitem}
\usepackage{mathtools}
\usepackage[utf8]{inputenc}
\usepackage{amssymb}
\usepackage{gensymb}

\usepackage[compact]{titlesec}
    \titlespacing{\section}{0pt}{2ex}{1ex}
    \titlespacing{\subsection}{0pt}{1ex}{0ex}
    \titlespacing{\subsubsection}{0pt}{0.5ex}{0ex}
\usepackage{float}
\usepackage{array}
\usepackage{algorithmic}
\usepackage{textcomp}
\usepackage{xcolor}
\usepackage{tabularx}
\def\BibTeX{{\rm B\kern-.05em{\sc i\kern-.025em b}\kern-.08em
    T\kern-.1667em\lower.7ex\hbox{E}\kern-.125emX}}

\begin{document}

\include{bibliography}

\title{Accurate Current Sharing in a DC Microgrid Using Modified Droop Control Algorithm\\
}

\author{
\IEEEauthorblockN{Naser Souri, \emph{Graduate Student Member, IEEE}, Ali Mehrizi-Sani\thanks{This work is supported in part by the National Science Foundation (NSF) under award ECCS-1953213, in part by the State of Virginia’s Commonwealth Cyber Initiative (www.cyberinitiative.org), in part by the U.S. Department of Energy’s Office of Energy Efficiency and Renewable Energy (EERE) under the Solar Energy Technologies Office Award Number 38637 (UNIFI Consortium led by NREL), and in part by Manitoba Hydro International. The views expressed herein do not necessarily represent the views of the U.S. Department of Energy or the United States Government.}, \emph{Senior Member, IEEE}}
\IEEEauthorblockA{The Bradley Department of Electrical and Computer Engineering 
Virginia Tech, Blacksburg, VA 24061 \\ \thanks{“© 20XX IEEE.  Personal use of this material is permitted.  Permission from IEEE must be obtained for all other uses, in any current or future media, including reprinting/republishing this material for advertising or promotional purposes, creating new collective works, for resale or redistribution to servers or lists, or reuse of any copyrighted component of this work in other works.”}
e-mails: \{nsouri,mehrizi\}@vt.edu} 

}

\maketitle

\begin{abstract}
Due to the increasing popularity of DC loads and the potential for higher efficiency, DC microgrids are gaining significant attention. DC microgrids utilize multiple parallel converters to deliver sufficient power to the load. However, a key challenge arises when connecting these converters to a common DC bus: maintaining voltage regulation and accurate current sharing. Unequal cable resistances can cause uneven power sharing and lead to power losses. Conventional droop control methods, which employ a virtual resistor to address this issue, have limitations in achieving good performance across the entire converter operating range. This paper proposes a modified droop control algorithm to address this issue. This method modifies the virtual resistor in a way that ensures power sharing aligns with each converter-rated capacity. The algorithm is simple to implement and uses local measurements to update the droop gain. This paper presents simulation studies and experimental tests to analyze the performance of the proposed method, considering scenarios with equal and unequal converter ratings. The results successfully validate the accuracy and effectiveness of this innovative approach.

\end{abstract}


\section{Introduction}
The global shift toward renewable energy sources such as solar and wind is driven by both environmental necessity and policies around the world \cite{Farhangi, Sirat }. Integrating these resources into power grids, potentially as a microgrid, offers a dual benefit: increased grid resiliency and reduced dependence on fossil fuels. Since many loads are powered by DC, eliminating the DC-to-AC conversion step inherent in AC microgrids potentially results in a more efficient system in DC microgrids \cite{souri2024stability}. 
However, the widespread adoption of DC microgrids requires addressing key technical challenges. These challenges encompass the design and control strategies needed for optimal operation, ensuring system stability, especially during dynamic changes, and developing robust protection schemes \cite{Ali,Kheirollahi}.

Paralleling DC/DC converters in a microgrid increases current capacity and improves system redundancy. However, the incorporation of parallel converters increases the complexity of microgrid control. In addition, they can pose these challenges: degradation in voltage regulation, unequal power sharing between converters, and the emergence of circulating currents \cite{Xiaonan, Mehrdad}. These issues contribute to a decrease in overall system efficiency. Furthermore, there is a chance that an unequal power distribution among parallel converters will cause them to exceed their rated voltage or current. 

Maintaining a balanced power distribution across interconnected systems is paramount. Extensive research efforts address this challenge in both AC and DC grids. The focus lies on developing effective droop-based power sharing strategies for the multiple converters, which can be found in \cite{Xiaochao, Jiahang, Derbas}. Reference \cite{Jiahang} considers the local load on the power sharing and tries to mitigate the issue using a virtual voltage instead of the actual output voltage. The virtual voltage is constructed using the average voltage of all converters. However, correcting the current is more important since the voltages at the converter terminal may differ from each other with different cable resistors. Hence, \cite{Derbas} proposes a strategy based on the average current of the converters. Using communication between the converters, the average current is compared with the converter output current, and a corrective term is added to the outer loop of the converter. However, unequal power sharing is not taken into account, and the dynamic behavior of the converter is focused on. 

Reference \cite{Baharizadeh} proposes a two-layer control strategy that, unlike the conventional method ($P-V$) is based on $P-\Dot{V}$ to remove the dependency of the sources on the line resistors. Additionally, the upper layer utilizes a secondary control for voltage compensation. However, this method includes a derivative term in the control loop and may amplify the noise and degrade the performance. 

Although conventional droop control exhibits poor performance over a wider range of operating points, nonlinear methods are introduced. These methods provide a higher droop gain as the current increases. Thus, the output current of these strategies is a function of the current. For example, \cite{Burgos} proposes a nonlinear method in which the droop resistor increases at a heavy load. Furthermore, \cite{Prabhakaran} proposes a nonlinear method based on a polynomial function for the droop gain. In this method, the droop gain changes based on a polynomial function to increase the power sharing accuracy. However, nonlinear methods exhibit better performance than conventional methods; they still lack accuracy as the line resistors are not known. Adoptive methods, therefore, are introduced to modify the droop gain online. Hence, \cite{Jung} proposes an adaptive droop control approach; however, it assumes that all power sources have identical capacities. 

Therefore, the contribution of this paper is:
\begin{itemize}
  \item Proposing an adaptive droop gain for equal or proportional power sharing between identical or different converters by sharing only the current through communication lines. 
\end{itemize} 
The following is the remainder of the paper.

Section II provides a system description and current sharing formulation and discusses the limitations of conventional droop control for parallel converters.
Section III delves into the proposed algorithm for current sharing and explains the method. Section IV provides simulation studies in the MATLAB/Simulink environment. In this section, three different cases are considered to evaluate the performance of the proposed method. The last section provides two experimental tests to validate the simulation studies.


\section{Parallel Converters and Problem Statement}
Ensuring balanced power distribution among these converters, known as power sharing, is crucial to optimal performance and system stability. This section explores the concept of parallel DC/DC converters and the associated challenge of current sharing. 

\subsection{System Configuration}

\begin{figure}[!t]
\centerline{\includegraphics[width= 0.85\columnwidth ]{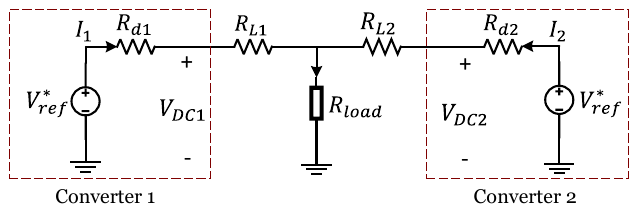}}

\vspace*{-0.4 cm}
\caption{A simple equivalent circuit for parallel DC/DCs in a DC microgrid.}
\label{fig:DC_microgrid}

\vspace*{-0.4 cm}
\end{figure}

A typical DC microgrid consists of various DERs, energy storage units, and loads connected to a common DC bus. Each DER may require a dedicated DC/DC converter to interface with the DC bus. These DC/DC converters are in parallel to deliver sufficient current to the load. Fig.~\ref{fig:DC_microgrid} shows a simplified DC microgrid system. The system consists of two parallel connected converters supplying the load. $R_{ L1 }$ and $R_{ L2 }$ are the cable resistors, and $R_{ d1 }$ and $R_{ d2 }$ are the droop gains, which are explained later in detail. $V_{DC1}$ and $V_{DC2}$ represent the converter terminal voltages. $I_1$ and $I_2$ represent the converter output current. $V^*_{ref}$ represents the reference voltages for the converters. 

\begin{figure*}[htbp]
\centerline{\includegraphics[width=1.668\columnwidth ]{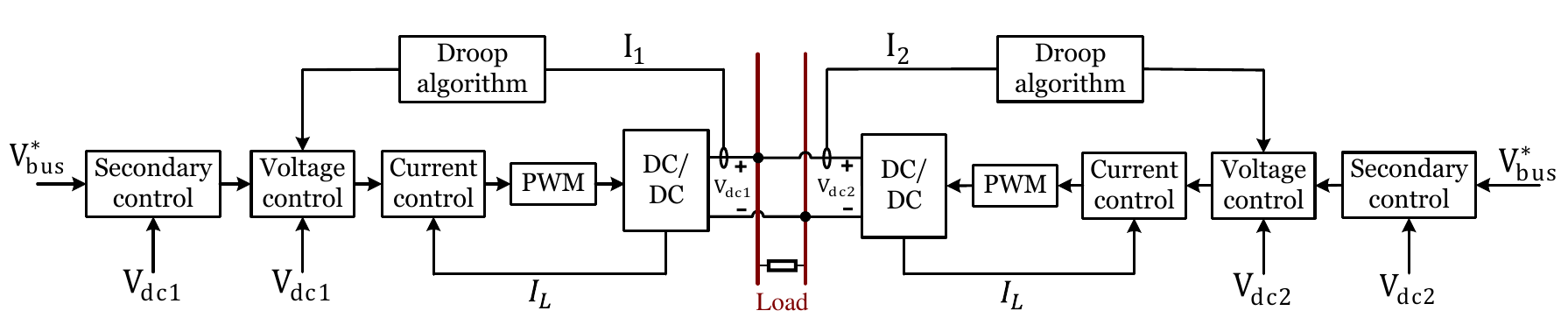}}

\vspace*{-0.4 cm}
\caption{General control block diagram of parallel converters.}
\label{fig:general_control_diagram}

\vspace*{-0.4 cm}
\end{figure*}

\subsection{Current Sharing Analysis for Parallel Converters}

This section analyzes the parallel converters mathematically. Using Kirchhoff’s laws for the circuit shown in Fig.~\ref{fig:DC_microgrid} and neglecting the droop gains, the output current of the converters is obtained as follows:

\begin{equation}
I_1 = \frac{R_2V_1 +R_L(V_{DC1} - V_{DC2})}{R_LR_{L1} + R_LR_{L2} + R_{L1}R_{L2}}.
\end{equation}

\begin{equation}
I_2 = \frac{R_1V_2 +R_L(V_{DC2} - V_{DC1})}{R_LR_{L1} + R_LR_{L2} + R_{L1}R_{L2}}.
\end{equation}

Any differences in the converter DC voltages and cable resistances lead to uneven current sharing between the converters. The more the cable resistors are different, the larger the current difference becomes. This difference between the converters is obtained from (1) and (2):

\begin{equation}
\begin{split}
& \Delta I = I_1 - I_2 = \\
& \frac{ 2R_L(V_{DC1} - V_{DC2}) }{R_LR_{L1} + R_LR_{L2} + R_{L1}R_{L2}} + \frac{R_2V_{DC1} - R1V_{DC2} }{R_LR_{L1} + R_LR_{L2} + R_{L1}R_{L2}}.
\end{split}
\end{equation}

To share an equal current of each converter, two conditions should be considered:

\begin{equation*}
V_{DC1} = V_{DC2}  \quad  \text{and} \quad R_{L1} = R_{L2}.
\end{equation*}

However, in practice, achieving an equal amount of resistors for the cables is impossible. Hence, there is always a difference in the current. Droop control is introduced to overcome this issue, which is discussed in the next section.

\subsection{Conventional Droop Control Limitations and Solution}

To reduce the current difference between the converters, a droop gain is introduced, in which a virtual resistor is added in series to the cable resistor to compensate for the difference in the cable resistances, as shown in Fig.~\ref{fig:DC_microgrid}. 
Considering the same value of voltages for the converters, and adding virtual resistors, (3) is updated as follows:

\begin{equation}
\Delta I=  \frac{R'_{L2}V_{DC1} - R'_{L1}V_{DC2} }{R_LR'_{L1} + R_LR'_{L2} + R'_{L1}R'_{L2}}.
\end{equation}

\noindent where the virtual resistors are considered in $R'_{L1} = R_{L1} + R_{d1}$ and $R'_{L2} = R_{L2} + R_{d2}$. According to (4), to reduce the current difference, the following condition should be considered in the control loop of each converter:

\begin{equation}
\frac{R'_{L1}}{R'_{L2}} = \frac{V_{DC1}}{V_{DC2}}.
\end{equation}

\begin{figure}[!t]
\centerline{\includegraphics[width= 0.95\columnwidth ]{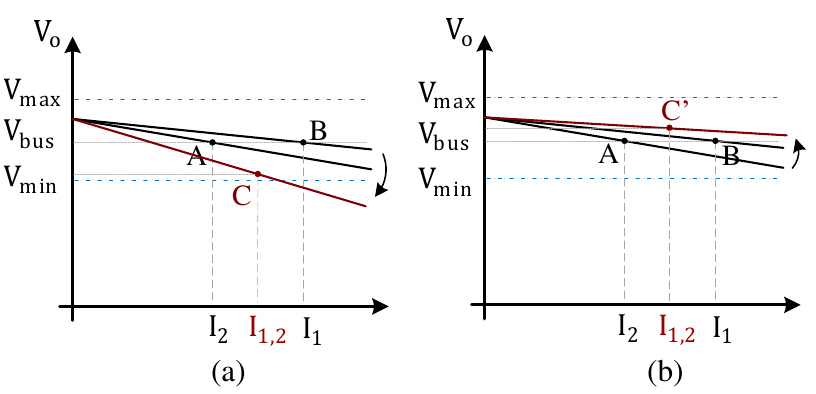}}

\vspace*{-0.4 cm}
\caption{V-I characteristic of the converters with and without using droop control.}

\vspace*{-0.4 cm}
\label{fig:V_I_characteristic}
\end{figure}

Therefore, to share the same current and mitigate the current difference, the total line resistors should match using the droop control. Fig.~\ref{fig:V_I_characteristic} shows the voltage-current (V-I) characteristic for the parallel converters before and after adding virtual resistors. Before adding the droop resistors, the difference in the converter current is large, and it gets smaller when the droop resistors are added to the control loop. Although conventional droop control offers a simple and effective way to manage power sharing in DC microgrids, it is not without limitations. These limitations can manifest as challenges in areas such as power-sharing accuracy in the entire range of converter operations. Additionally, droop control can lead to voltage regulation issues. These limitations become more pronounced when renewable energy sources are integrated with their inherent variability. As a result, researchers are actively exploring more sophisticated control strategies to address these shortcomings. 
Therefore, in the next section, a modified droop control method is proposed to overcome conventional droop control.

\section{Modified Droop Control for Power Sharing}
\begin{figure}
\centerline{\includegraphics[width= 0.7\columnwidth ]{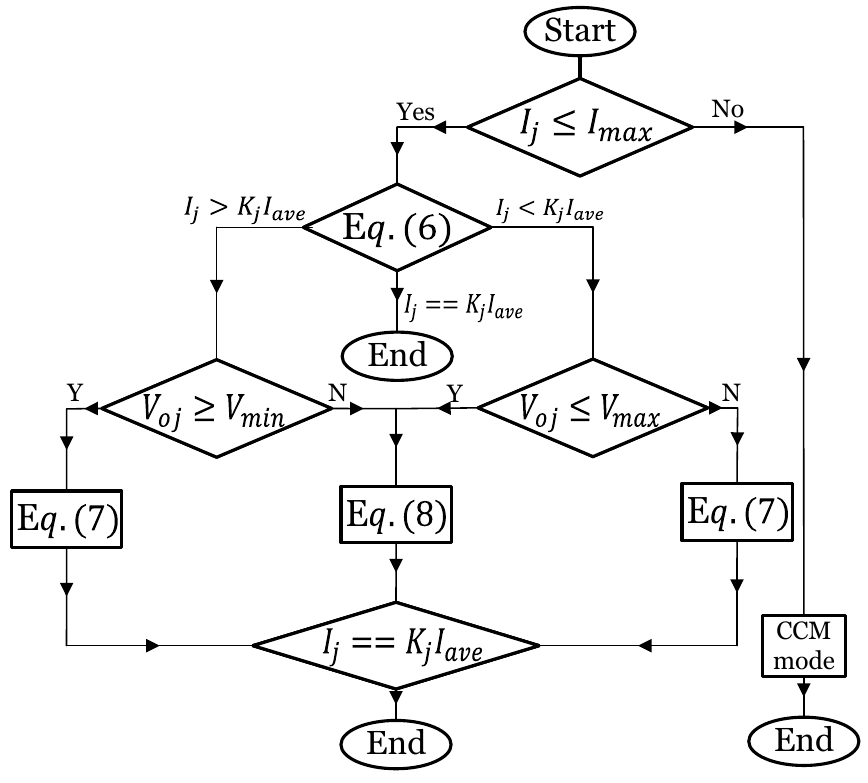}}

\vspace*{-0.4 cm}
\caption{The proposed algorithm for power sharing.}

\vspace*{-0.4 cm}
\label{fig:droop_algorithm}
\end{figure}

This section discusses the proposed method for power sharing in a DC microgrid. The droop resistors are constant in the conventional droop control. A small droop gain results in a current difference, and a big droop gain leads to voltage regulation issues. Therefore, this paper utilizes local measurements to provide an appropriate droop gain that is adaptive to changes in the grid or unknown cable resistors. This algorithm is demonstrated in Fig.~\ref{fig:droop_algorithm}. By monitoring the local voltage and current of the converter and comparing it with the provided reference, power sharing is achieved. This reference can be obtained from the average current of the converters times the load factor:

\begin{equation}
|I_j - K_jI_{ave}|<\gamma.
\end{equation}

\noindent where $I_j$ is the converter current, and $K_j$ is a coefficient, which is the load factor (e.g., $K_j=0.8$ means the converter provides 80\% of its rated power for the load.). $\gamma$ is a small value that represents the accuracy of current sharing.
The converter current can be transferred through low bandwidth communication, as smart grids utilize a combination of communication technologies, allowing remote monitoring \cite{Tariq}. 
By comparing the average current with the local current, the controller decides whether to increase or decrease the droop gain and modifies the voltage reference:

\begin{equation}
R_d(j) = R_{d int} + R_{dold} + \delta.
\end{equation}
\begin{equation}
R_d(j) = R_{d int} + R_{dold} - \delta.
\end{equation}

\noindent where $R_d(j)$ is the droop resistor for the converter $j^{th}$, $R_{d int}$ is the initial droop resistor, and $R_{dold}$ is a parameter to hold the previous gain in the control loop. $\delta$ represents the step change to update the droop gain. This value determines the speed with which the droop gains are updated. Before updating the droop gain, the converter voltage is monitored so that it does not exceed the normal operation. In addition, if the converter current exceeds the nominal current, the algorithm is stopped and enters the current control mode (CCM). The general control block diagram of the converters is shown in Fig.~\ref{fig:general_control_diagram}. The control block diagram uses a cascaded PI controller for the voltage and current. A secondary PI controller is added to the voltage control to compensate for the bus voltage deviation. Hence, the voltage from point (c) changes to point (c') as shown in Fig.~\ref{fig:V_I_characteristic}(b).

\section{Simulation Results}
The effectiveness of the proposed method is assessed through simulations developed in MATLAB/Simulink environment. These simulation studies employed the same system parameters detailed in Table I. These studies examine two parallel buck converters that are connected to the load through different line resistors. Three case studies are considered to evaluate the proposed method for active damping.

\subsection{Case I: Droop Control Activation After a While}
\begin{figure}
\centering
\includegraphics[width=0.99\columnwidth]{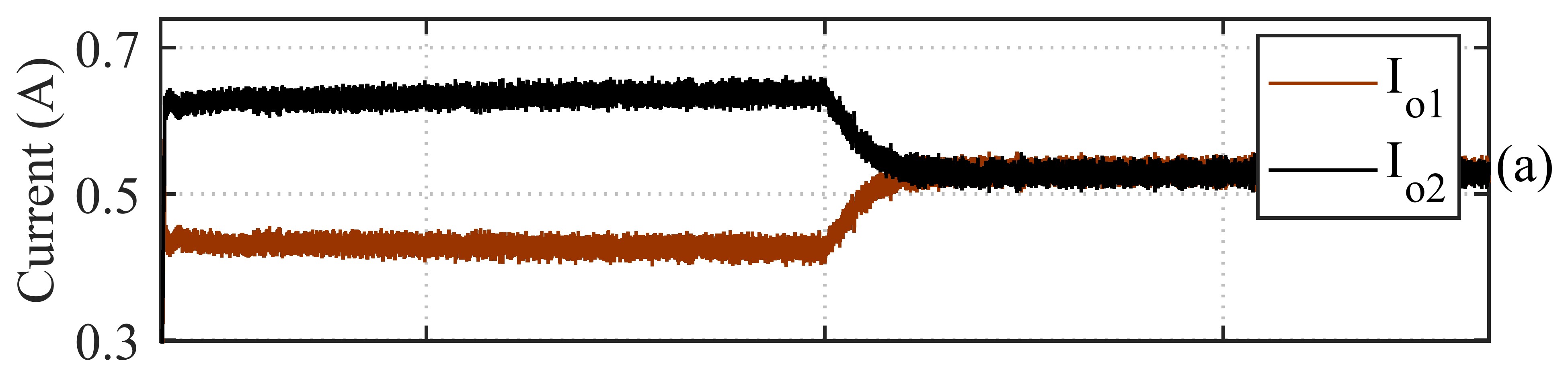}

\vspace*{-0.0 cm}
\hfill
\includegraphics[width=0.99\columnwidth]{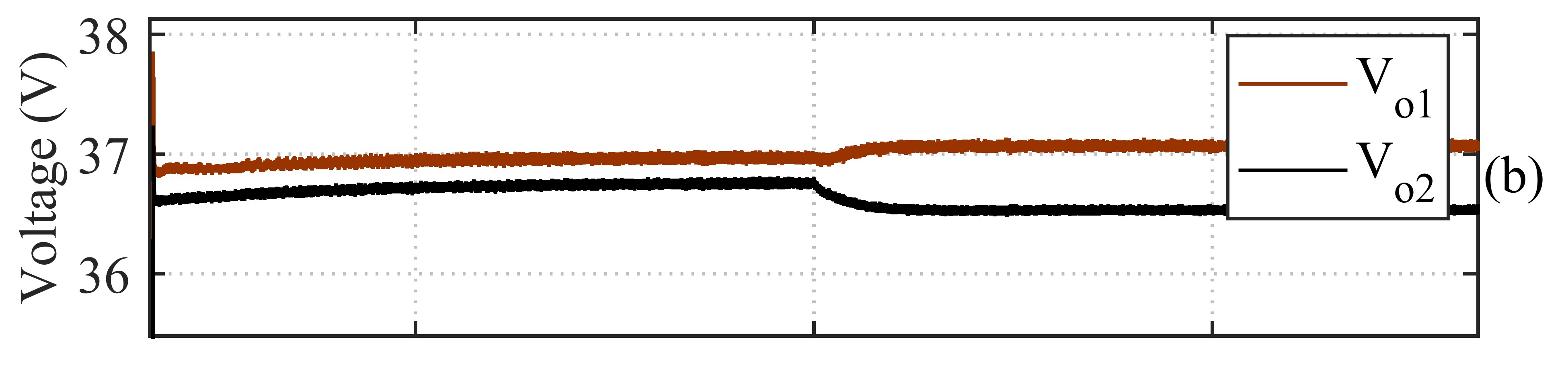}

\vspace*{-0.0 cm}
\hfill
\includegraphics[width=0.99\columnwidth]{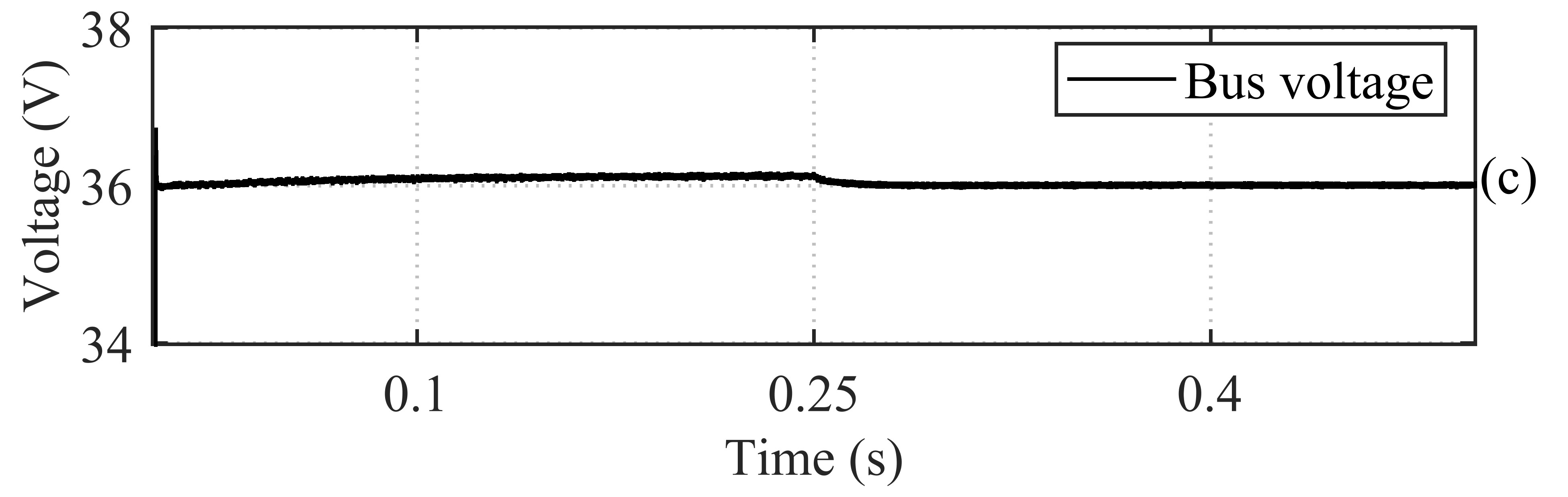}

\vspace*{-0.45 cm}
\caption{Case study I: voltages and currents of the converters before and after activation of droop control with $R_{L1}=1~\Omega$ and $R_{L1}=2~\Omega$: (a) converter current, (b) terminal voltages, (c) bus voltage.}

\vspace*{-0.3 cm}

\label{fig:case1}
\end{figure}

In this case study, the converters are in parallel with different cable resistors. First, the droop gains are zero, and then the proposed droop control is activated at $t=0.25$~ms. In this case, the algorithm looks for an appropriate droop gain to share the same amount of power. The result of this case study, which is shown in Fig.~\ref{fig:case1}, shows that the converters share the same current quickly. The converter voltages, which are shown in Fig.~\ref{fig:case1}(b), change slightly to correct for the injected power to the load. The bus voltage deviation is in an acceptable range. Fig.~\ref{fig:case1}(c), which is the bus voltage, has small changes after power sharing activation.

\subsection{Case II: Load Factor Changing }

\begin{figure}
\centering
\includegraphics[width=0.99\columnwidth]{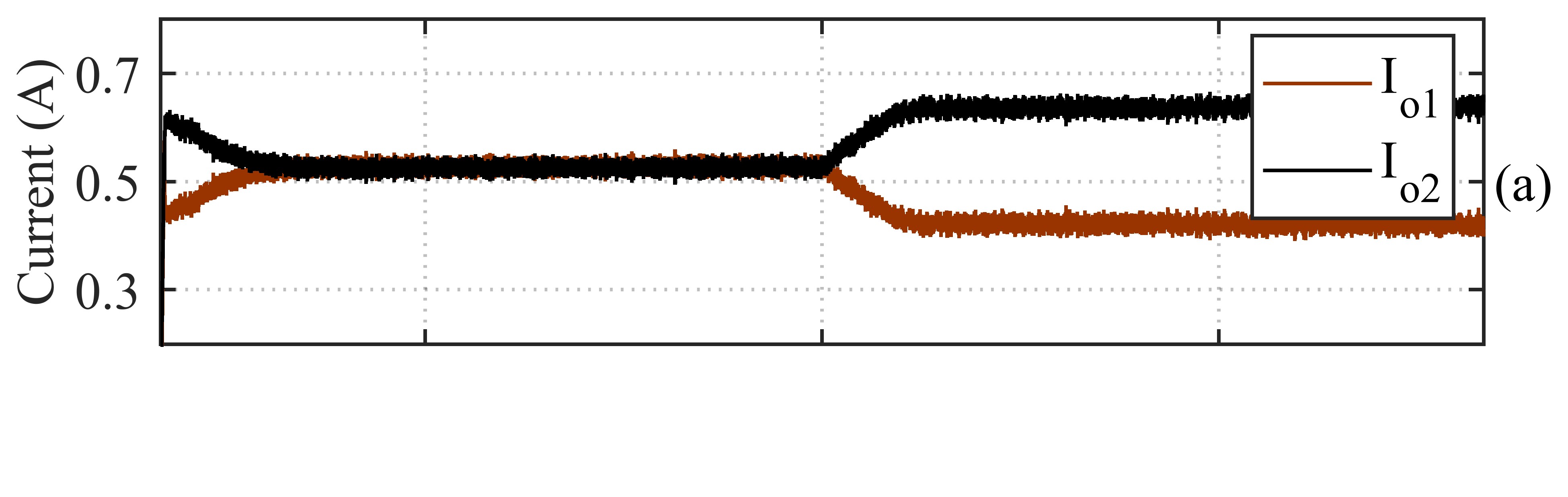}

\vspace*{-0.65 cm}
\hfill
\includegraphics[width=0.99\columnwidth]{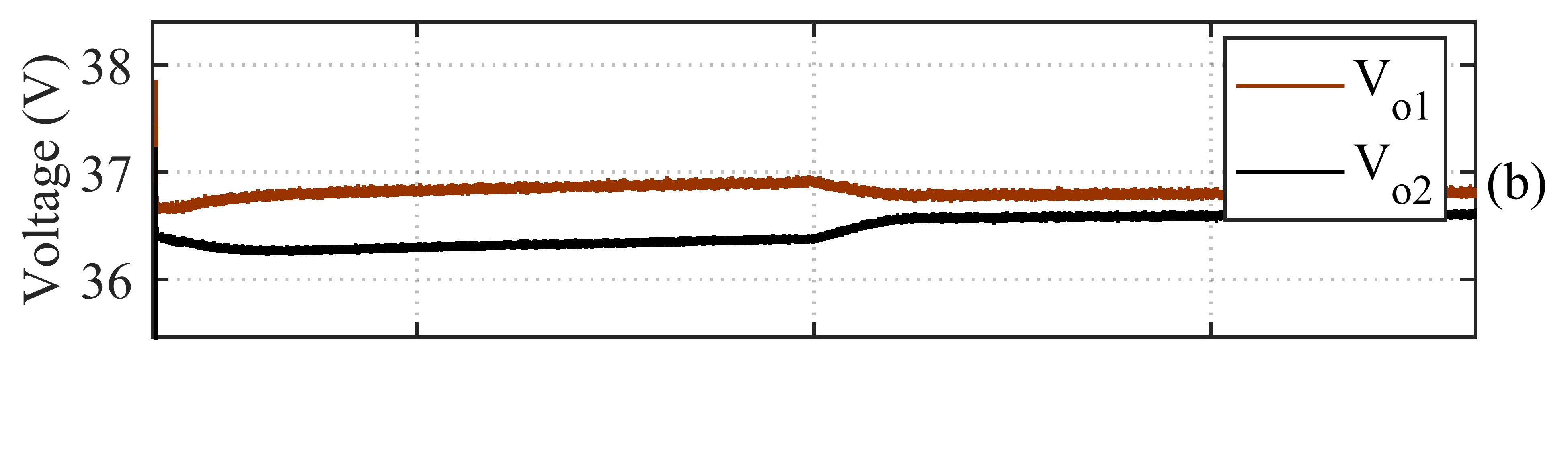}

\vspace*{-0.5 cm}
\hfill
\includegraphics[width=0.99\columnwidth]{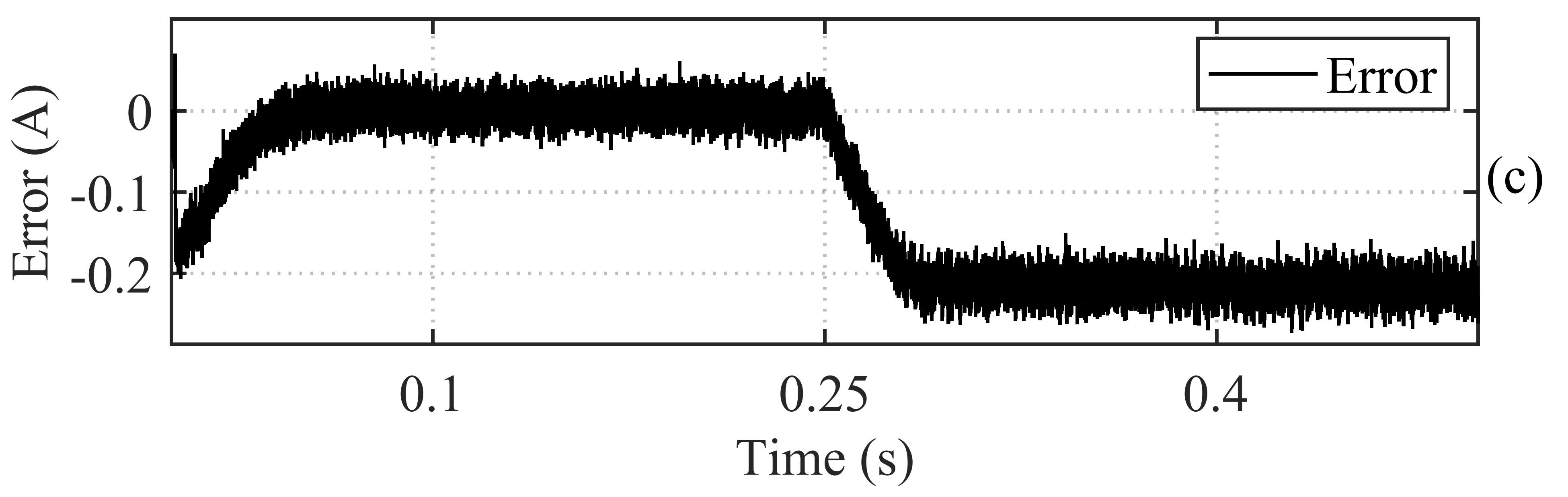}

\vspace*{-0.45 cm}
\caption{Case study II: voltages and currents with changing the load factor with $R_{L1}=1~\Omega$ and $R_{L1}=2~\Omega$: (a) converter current, (b) terminal voltages, (c) the current error between the converters.}

\vspace*{-0.4 cm}

\label{fig:case2}
\end{figure}

In the second case study, two parallel DC/DC converters, same as in case I, start with the droop control. Hence, they share the same current. After a while, the converters sharing power amount changes. By changing the load factor for the converters at $t=0.25$~ms, one of them provides more power for the load, and the other one provides the remaining power. The result for current and voltages in this case is shown in Fig.~\ref{fig:case2}. The difference in current between the converters is shown in Fig.~\ref{fig:case2}(c), which validates the accuracy and performance of the proposed method.

\begin{figure}
\centering
\hfill
\includegraphics[width=0.99\columnwidth]{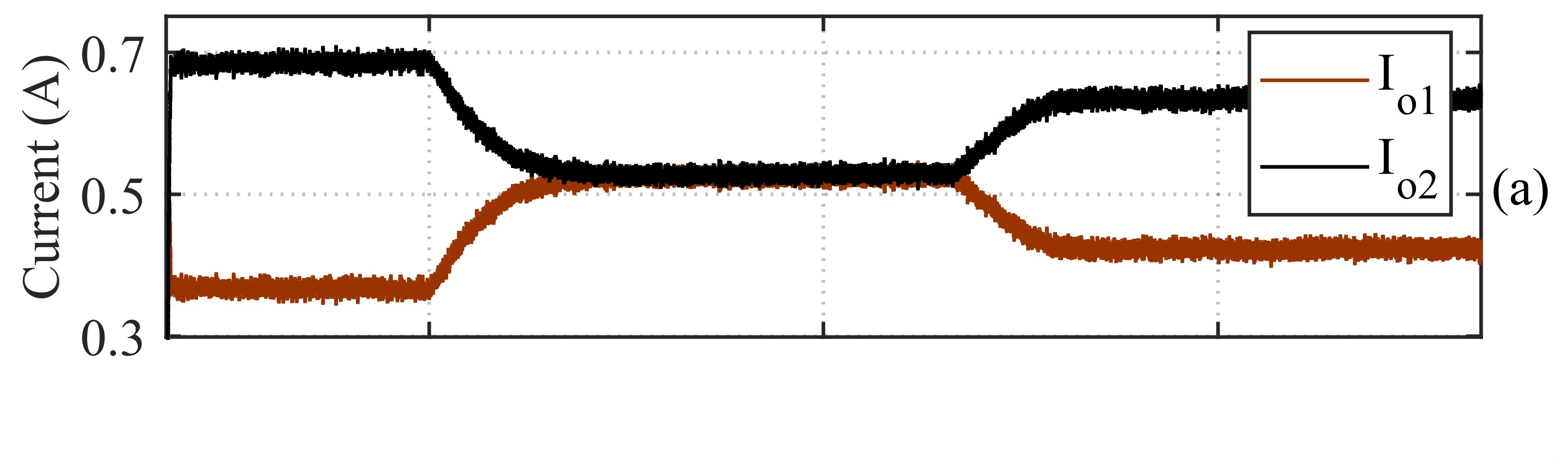}


\vspace*{-0.66 cm}
\hfill
\includegraphics[width=0.99\columnwidth]{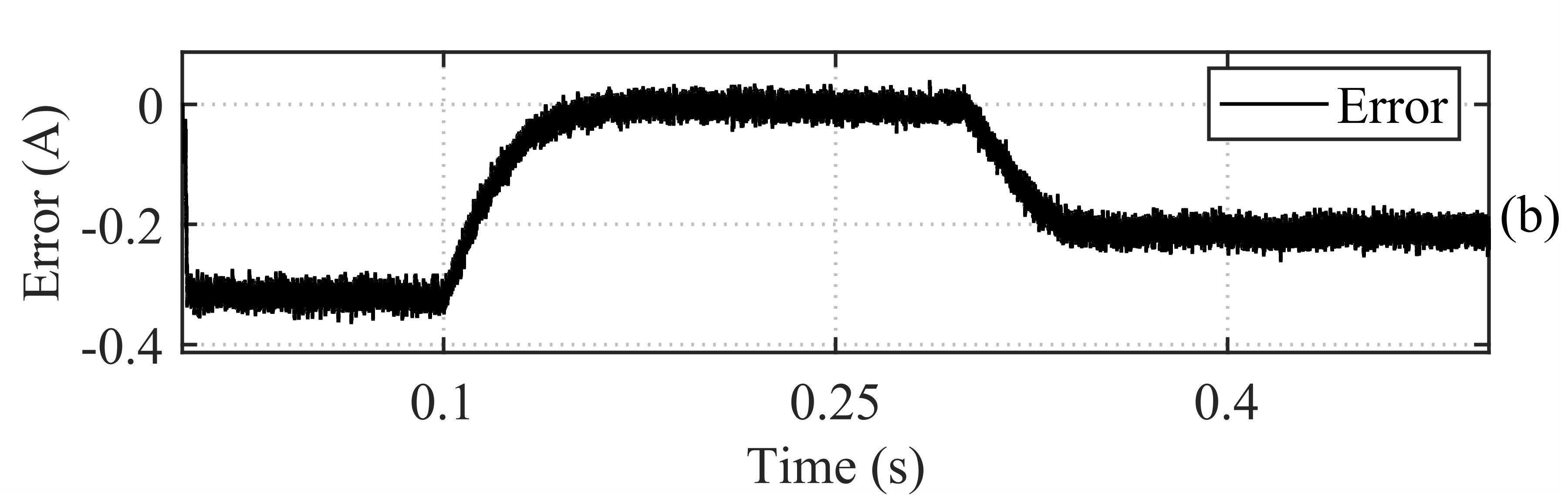}

\vspace*{-0.43 cm}
\caption{Case study III: voltages and currents of the converters before and after activation of droop control and load factor changing with $R_{L1}=1~ \Omega $ and $R_{L1}=3~\Omega$: (a) converter current, (b) current error.}
\label{fig:case3}
\end{figure}

\subsection{Case III: Changing Cable Resistances }
In this case study, the system is the same as in previous case studies, but the cable resistor changes. It is assumed that one of the converters is located far from the other. Therefore, the cable resistor changes from $2~\Omega$ to $3~\Omega$. The simulation starts without droop control. It is activated at $t=0.1$~ms, and the load factor changes at $t=0.3$~ms. The result for the current is shown in Fig.~\ref{fig:case3}(a). Furthermore, the current error between the converters is shown in Fig.~\ref{fig:case3}(b). When the converters share the same current, this error gets zero.

\begin{table}[t]
\caption{System Parameters}

\vspace*{-0.4 cm}
\begin{center}
\begin{tabular}{lcclcc}

\toprule
\textbf{Parameter} & \textbf{Symbol} & \textbf{Value}  & \textbf{Parameter}  & \textbf{Symbol} & \textbf{Value} \\ 
\toprule
\toprule

\begin{tabular}[c]{@{}l@{}}Input \\ voltage\end{tabular}         & $V_g$              & \begin{tabular}[c]{@{}l@{}}60 V\\ \end{tabular} & \begin{tabular}[c]{@{}l@{}}Output\\ voltage \end{tabular}    & $V_O$              & 36 V         \\ 
\midrule
\begin{tabular}[c]{@{}l@{}}Cable 1 \\ resistor\end{tabular}      & $R_{L1}$             & 1 $\Omega$                                                 & \begin{tabular}[c]{@{}l@{}}Cable 2\\ resistor\end{tabular} & $R_{L2}$              & 2 $\Omega$         \\ 
\midrule
\begin{tabular}[c]{@{}l@{}}Rated \\ power \end{tabular} & $P_{out}$              & 100 W                                                 & \begin{tabular}[c]{@{}l@{}} Inductor \\ \end{tabular}        & $L_{buck}$              &  2 mH         \\ 
\midrule
\begin{tabular}[c]{@{}l@{}}Capacitor\\  \end{tabular}   & $C_{buck}$             & 100 uF                                                & \begin{tabular}[c]{@{}l@{}}Switching\\ frequency\end{tabular}   & $F_s$              & 25 kHz        \\ 

\midrule
\begin{tabular}[c]{@{}l@{}}Empirical  \\ line resistor   \end{tabular}   & $R_{L1}$             & 2 $\Omega$                                                & \begin{tabular}[c]{@{}l@{}}Empirical  \\ line resistor \end{tabular}   &  $R_{L2}$              & 4 $\Omega$        \\ 
\bottomrule
\end{tabular}
\label{table:system_parameters}
\end{center}

\vspace*{-0.65 cm}
\end{table}

\section{Experimental Test}

\begin{figure}[t]
\centerline{\includegraphics[width= .94\columnwidth ]{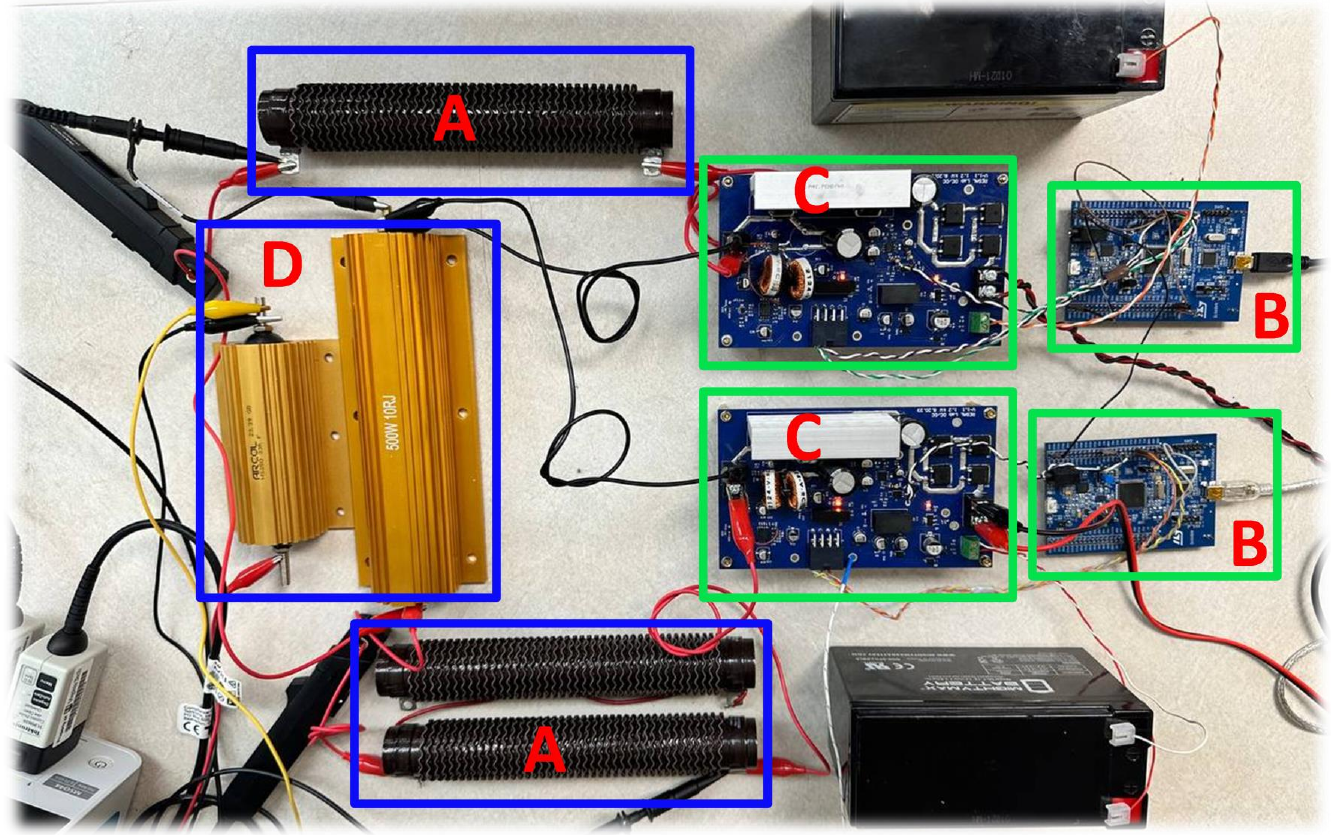}}
\vspace*{-0.2 cm} 
\caption{Experimental setup of DC microgrid containing parallel converters.}
\label{fig:Experimental_setup}

\vspace*{-0.1 cm}
\end{figure}

To verify the effectiveness of the proposed method for power sharing, a scaled-down prototype microgrid is constructed in the lab. This prototype consists of two parallel buck converters with a power of 100~W. The system parameters used in this prototype mirror those presented in Table I, which are the same as those used in the simulation study in MATLAB. The experimental setup is shown in Fig.~\ref{fig:Experimental_setup}. In this figure, (A) is the cable resistors, (B) is the converter controller, (C) is the DC/DC converters, and (D) is the load. By analyzing the output current and voltage waveforms, we were able to evaluate the performance of the control system. 

\subsection{Power Sharing Activation With Different Line Resistors}
\begin{figure}[t]
\centerline{\includegraphics[width= 0.9\columnwidth ]{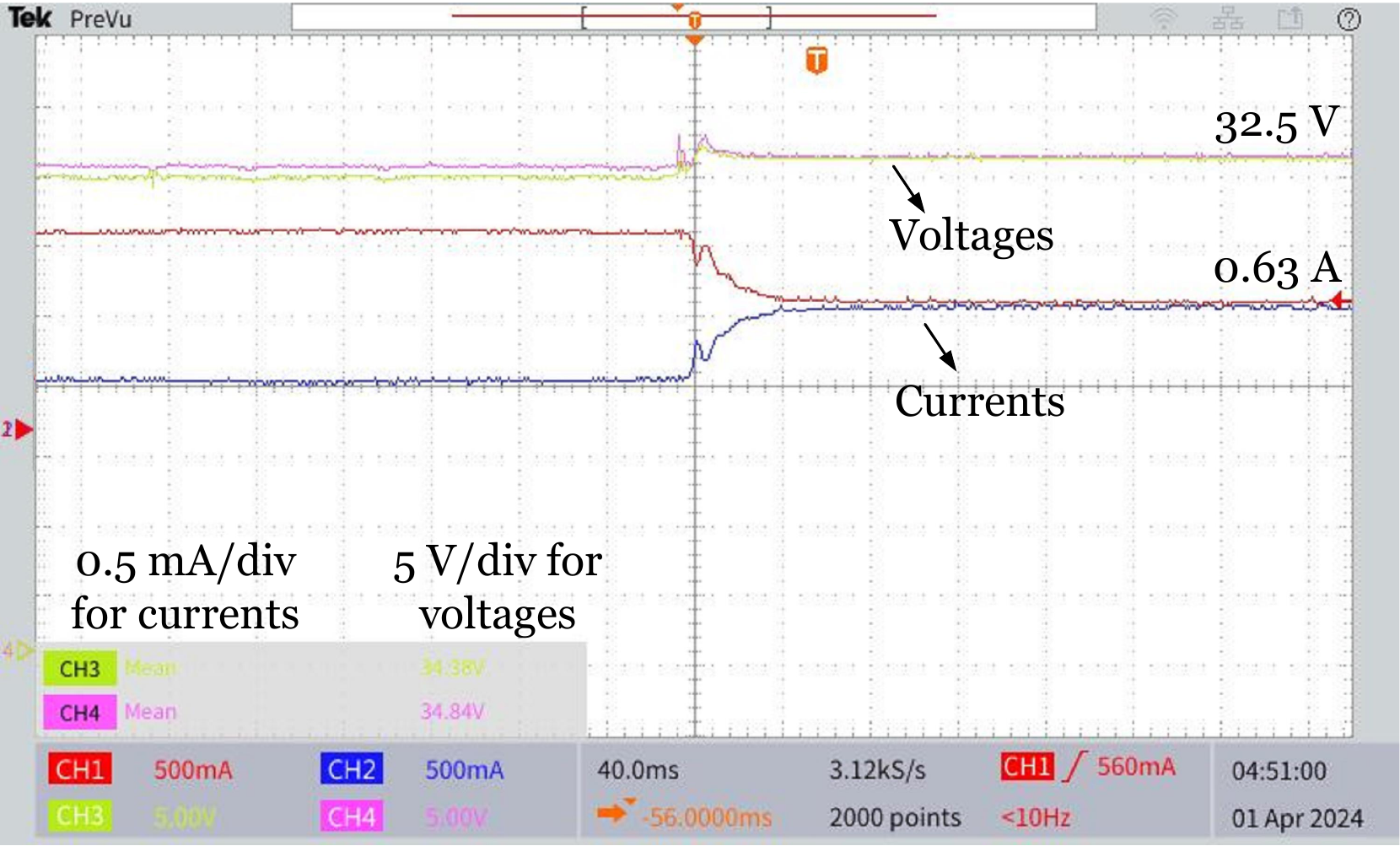}}
\vspace*{-0.2 cm} 
\caption{Currents and voltages of the converters after activation of the proposed droop control in the experimental test.}

\vspace*{-0.2 cm}
\label{fig:power_sharing_activation}

\vspace*{-0. cm}
\end{figure}

In this case, the converters are in parallel with different line resistors to supply the load. At first, the proposed algorithm is deactivated. The line resistors for the converters are $R_{L1}=2~\Omega$ and $R_{L2}=4~\Omega$. Thus, they share an unequal power. After a while, the droop control is activated, and they share the same current as shown in Fig.~\ref{fig:power_sharing_activation}. 

\subsection{Proportional Power Sharing and Step Change in the Load}
\begin{figure}[t]
\centerline{\includegraphics[width= 0.9\columnwidth ]{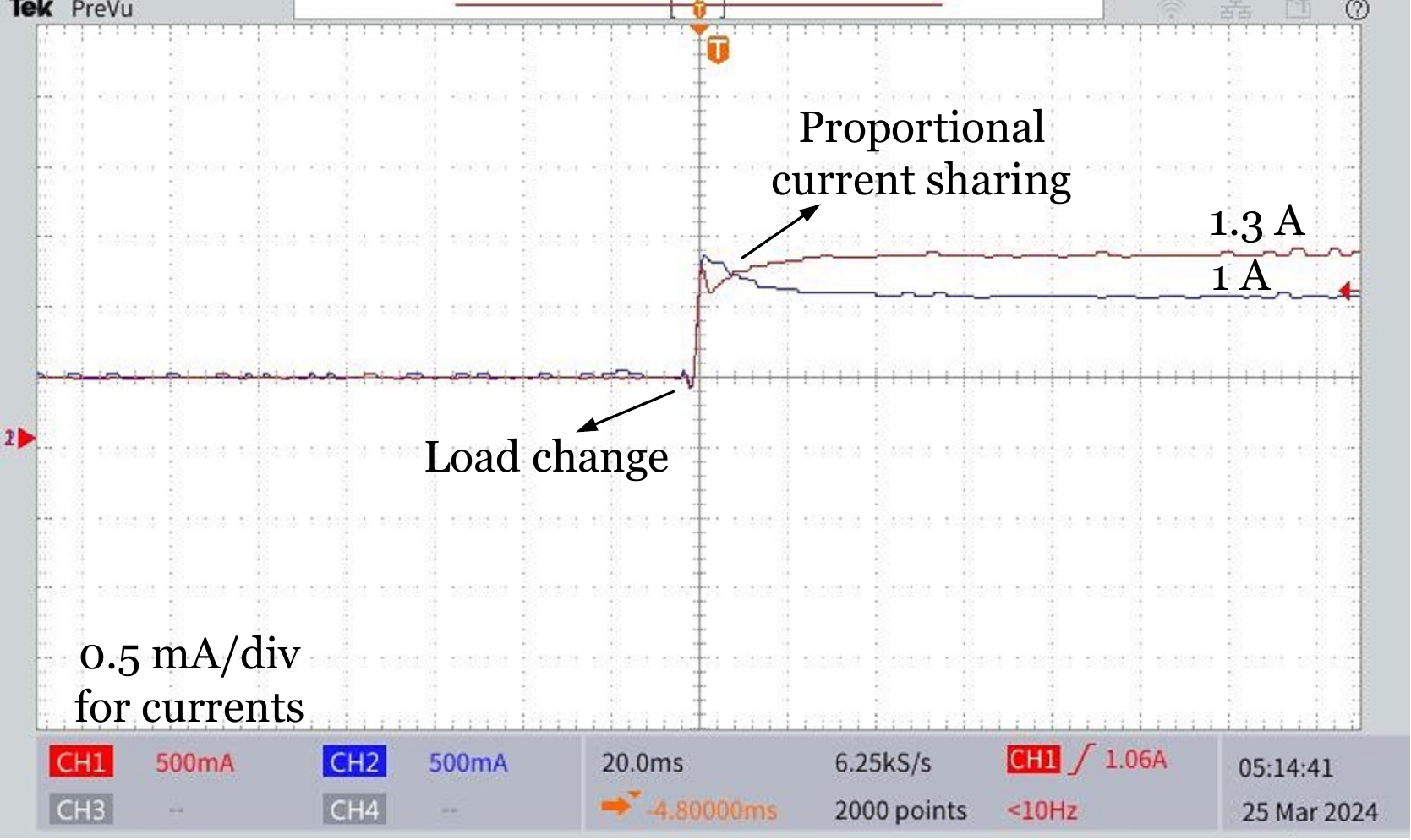}}
\vspace*{-0.0 cm} 
\caption{Proportional current sharing waveform with a step change in the load.}

\vspace*{-0.5 cm}
\label{fig:step_change_proportinal}

\vspace*{-0.0 cm}
\end{figure}

In this case, the converters share the same current, and after a while, a step change in the load occurs. Demanding more power from the converters, one of them shares more power for the load, according to the LF. The results are shown in Fig.~\ref{fig:step_change_proportinal}, which validate the proposed method for different case studies.

\section{Conclusion}
This work addresses limitations in droop control for DC microgrids by proposing a modified droop method. The proposed method dynamically adjusts droop gain via online calculations, resulting in a computationally efficient algorithm. In addition, the method offers flexibility in handling unequal power distribution scenarios in which sources may have limited power output. This method relies on communication lines. The impact of line resistors on performance can be used in future research. Verification is done through a simulation in MATLAB/Simulink and an experimental setup. The results verify the performance and accuracy of the proposed method.

\bibliographystyle{IEEEtran}
\bibliography{ref.bib}

\begin{thebibliography}{10}
\providecommand{\url}[1]{#1}
\csname url@samestyle\endcsname
\providecommand{\newblock}{\relax}
\providecommand{\bibinfo}[2]{#2}
\providecommand{\BIBentrySTDinterwordspacing}{\spaceskip=0pt\relax}
\providecommand{\BIBentryALTinterwordstretchfactor}{4}
\providecommand{\BIBentryALTinterwordspacing}{\spaceskip=\fontdimen2\font plus
\BIBentryALTinterwordstretchfactor\fontdimen3\font minus \fontdimen4\font\relax}
\providecommand{\BIBforeignlanguage}[2]{{%
\expandafter\ifx\csname l@#1\endcsname\relax
\typeout{** WARNING: IEEEtran.bst: No hyphenation pattern has been}%
\typeout{** loaded for the language `#1'. Using the pattern for}%
\typeout{** the default language instead.}%
\else
\language=\csname l@#1\endcsname
\fi
#2}}
\providecommand{\BIBdecl}{\relax}
\BIBdecl

\bibitem{Farhangi}
N.~Souri, S.~Farhangi, H.~Iman-Eini, and H.~Ziar, ``Modeling and estimation of the maximum power of solar arrays under partial shading conditions,'' in \emph{11th Power Electronics, Drive Systems, and Technologies Conference (PEDSTC), Tehran, Iran}, February 2020.

\bibitem{Sirat}
R.~A. Ghaderloo, A.~P. Sirat, and A.~Shoulaie, ``A high frequency active clamp forward converter with coreless transformer,'' in \emph{North American Power Symposium (NAPS), Asheville, NC, USA}, July 2023, pp. 1--6.

\bibitem{souri2024stability}
N.~Souri, A.~Mehrizi-Sani, and K.~Tehrani, ``Stability enhancement of {LCL}-type grid-following inverters using capacitor voltage active damping,'' in \emph{IEEE PES General Meeting, Seattle, WA, USA}, April 2024.

\bibitem{Ali}
D.~E. Olivares, A.~Mehrizi-Sani, A.~H. Etemadi, C.~A. Cañizares, R.~Iravani, M.~Kazerani, A.~H. Hajimiragha, O.~Gomis-Bellmunt, M.~Saeedifard, R.~Palma-Behnke, G.~A. Jiménez-Estévez, and N.~D. Hatziargyriou, ``Trends in microgrid control,'' \emph{IEEE Trans Smart Grid}, vol.~5, no.~4, pp. 1905--1919, May 2014.

\bibitem{Kheirollahi}
R.~Kheirollahi, S.~Zhao, S.~SalehiRad, A.~Mostafa, Z.~Zheng, H.~Zhang, and F.~Lu, ``Coordination of solid-state circuit breakers in multi-source {DC} microgrids using inverse time-current characteristic profile,'' in \emph{IEEE Energy Conversion Congress and Exposition (ECCE), Nashville, TN, USA}, October 2023, pp. 843--847.

\bibitem{Xiaonan}
X.~Lu, J.~M. Guerrero, K.~Sun, and J.~C. Vasquez, ``An improved droop control method for dc microgrids based on low bandwidth communication with {DC} bus voltage restoration and enhanced current sharing accuracy,'' \emph{IEEE Transactions on Power Electronics}, vol.~29, no.~4, pp. 1800--1812, June 2014.

\bibitem{Mehrdad}
M.~Yazdanian and A.~Mehrizi-Sani, ``Distributed control techniques in microgrids,'' \emph{IEEE Transactions on Smart Grid}, vol.~5, no.~6, pp. 2901--2909, April 2014.

\bibitem{Xiaochao}
Y.~Sun, X.~Hou, J.~Yang, H.~Han, M.~Su, and J.~M. Guerrero, ``New perspectives on droop control in {AC} microgrid,'' \emph{IEEE Transactions on Industrial Electronics}, vol.~64, no.~7, pp. 5741--5745, March 2017.

\bibitem{Jiahang}
\BIBentryALTinterwordspacing
Y.~Mi, J.~Guo, S.~Yu, P.~Cai, L.~Ji, Y.~Wang, D.~Yue, Y.~Fu, and C.~Jin, ``A power sharing strategy for islanded dc microgrid with unmatched line impedance and local load,'' \emph{Electric Power Systems Research}, vol. 192, p. 106983, December 2021. [Online]. Available: \url{https://www.sciencedirect.com/science/article/pii/S0378779620307811}
\BIBentrySTDinterwordspacing

\bibitem{Derbas}
A.~A. Derbas, M.~Kheradmandi, M.~Hamzeh, and N.~D. Hatziargyriou, ``A hybrid power sharing control to enhance the small signal stability in {DC} microgrids,'' \emph{IEEE Transactions on Smart Grid}, vol.~13, no.~3, pp. 1826--1837, March 2022.

\bibitem{Baharizadeh}
M.~Baharizadeh, M.~S. Golsorkhi, M.~Shahparasti, and M.~Savaghebi, ``A two-layer control scheme based on {P-V} droop characteristic for accurate power sharing and voltage regulation in {DC} microgrids,'' \emph{IEEE Transactions on Smart Grid}, pp. 1--1, February 2021.

\bibitem{Burgos}
F.~Chen, R.~Burgos, D.~Boroyevich, and W.~Zhang, ``A nonlinear droop method to improve voltage regulation and load sharing in dc systems,'' in \emph{2015 IEEE First International Conference on {DC} Microgrids {(ICDCM)}, Atlanta, GA, USA}, July 2015, pp. 45--50.

\bibitem{Prabhakaran}
P.~Prabhakaran, Y.~Goyal, and V.~Agarwal, ``Novel nonlinear droop control techniques to overcome the load sharing and voltage regulation issues in {DC} microgrid,'' \emph{IEEE Transactions on Power Electronics}, vol.~33, no.~5, pp. 4477--4487, September 2018.

\bibitem{Jung}
J.-W. Kim, H.-S. Choi, and B.~H. Cho, ``A novel droop method for converter parallel operation,'' \emph{IEEE Transactions on Power Electronics}, vol.~17, no.~1, pp. 25--32, january 2002.

\bibitem{Tariq}
M.~A. Taher, M.~Tariq, M.~Behnamfar, and A.~I. Sarwat, ``Analyzing replay attack impact in {DC} microgrid consensus control: Detection and mitigation by kalman-filter-based observer,'' \emph{IEEE Access}, vol.~11, pp. 121\,368--121\,378, September 2023.

\end{thebibliography}


\end{document}